\documentclass[reprint,
 amsmath,amssymb,aps,prl,footinbib]{revtex4-2}
\usepackage{times}
\usepackage{graphicx}
\usepackage{dcolumn}
\usepackage{bm}
\usepackage{mathtools}
\usepackage[english]{babel}
\usepackage{physics}
\usepackage{enumitem}
\usepackage{mathrsfs}
\usepackage{xfrac}
\usepackage{xr-hyper}
\usepackage{hyperref}
\usepackage{footmisc}
\usepackage{hypcap}
\usepackage{makecell}
\usepackage{url}
\usepackage{ifsym}
\usepackage{bbold}
\usepackage{gensymb}
\usepackage{multirow}
\usepackage{array}
\usepackage{longtable}
\usepackage{siunitx}
\usepackage{nicefrac}
\usepackage[dvipsnames]{xcolor}
\usepackage{ulem}

\newcommand{\iac}[1]{{\color{blue} #1 }}

\begin{document}

\title{Quantum vacuum excitation of a quasi-normal mode in an analog model of black hole spacetime}

\author{M. J. Jacquet$^{1\dagger}$}
\author{L. Giacomelli$^{2}$}
\author{Q. Valnais$^1$}
\author{M. Joly$^{1}$}
\author{F. Claude$^1$}
\author{E. Giacobino$^1$}
\author{Q. Glorieux$^1$}
\author{I. Carusotto$^2$}
\author{A. Bramati$^1$}

\affiliation{
$^1$ Laboratoire Kastler Brossel, Sorbonne Universit\'{e}, CNRS, ENS-Universit\'{e} PSL, Coll\`{e}ge de France, Paris 75005, France\\
$^2$ INO-CNR BEC Center and Dipartimento di Fisica, Università di Trento, via Sommarive 14, I-38123 Povo, Trento, Italy\\
$^\dagger$ correspondance to maxime.jacquet@lkb.upmc.fr}

\begin{abstract}%
Vacuum quantum fluctuations near horizons are known to yield correlated emission by the Hawking effect.
We use a driven-dissipative quantum fluid of microcavity polaritons as an analog model of a quantum field theory on a black-hole spacetime and numerically calculate correlated emission.
We show that, in addition to the Hawking effect at the sonic horizon, quantum fluctuations may result in a sizeable stationary excitation of a quasi-normal mode of the field theory.
Observable signatures of the excitation of the quasi-normal mode are found in the spatial density fluctuations as well as in the spectrum of Hawking emission.
This suggests an intrinsic fluctuation-driven mechanism leading to the quantum excitation of quasi-normal modes on black hole spacetimes.
\end{abstract}

\maketitle
Quantum fluctuations of fields in the vicinity of the event horizon of black holes (BHs) are predicted to cause the emission of correlated waves by the Hawking effect (HE)~\cite{Hawking}: while Hawking radiation propagates away from the horizon to outer space, the partner radiation falls inside the horizon.
Since signaling from inside the horizon is impossible, only the out-going Hawking radiation is observable and quantum correlations with the in-falling partner waves cannot be accessed.

The HE may also be observed in the laboratory thanks to analogue gravity setups~\cite{barcelo_analogue_2011,jacquet_next_2020}, namely condensed matter or optical systems,  engineered in such a way that their collective excitations propagate on effectively curved spacetimes~\cite{Unruh,visser_acoustic_1998}.
This idea has been experimentally demonstrated in a variety of platforms~\cite{philbin_fiber-optical_2008,rousseaux_observation_2008,lahav_realization_2010,weinfurtner_measurement_2011,jaskula_acoustic_2012,Nguyen,torres_rotational_2017,eckel_rapidly_2018,vocke_rotating_2018,wittemer_phonon_2019,euve_scattering_2020,jacquet_polariton_2020,patrick_backreaction_2021}:
for example, a horizon for sound waves forms in a one-dimensional trans-sonic fluid where the flow velocity of the fluid exceeds the speed of sound.
Crucially, observation on both sides of the horizon is possible with analogue setups and the HE has been detected via density correlations between Hawking radiation and its partner~\cite{balbinot_nonlocal_2008,carusotto_numerical_2008} in experiments based on both classical~\cite{euve_observation_2016} and quantum fluids~\cite{munoz_de_nova_observation_2019}.

In this Letter, we make use of a specific realisation of an effective spacetime realised in a driven-dissipative quantum fluid of exciton-polaritons in a semiconductor microcavity~\cite{Solnyshkov,Gerace,Nguyen,Grissins} to push forward the theoretical study of quantum fluctuations in the vicinity of horizons.
In particular, we investigate a one-dimensional trans-sonic configuration where the spatial shape of the quantum fluid and the external potential are optimised together to maximize the strength of the HE~\cite{jacquet_analogue_2022}.
We find that the typical signatures of the HE are supplemented by new features in the spatial correlation of sound waves that evidence the coupling of propagating waves \iac{to} a localized mode living near the horizon, namely a quasi-normal mode (QNM) of the acoustic field.
On a curved spacetime with a horizon, QNMs are decaying solutions of the Klein-Gordon equation of the field satisfying purely in-(out-)going boundary conditions at the horizon (at infinity)~\cite{chandrasekhar_quasi-normal_1975,berti_quasinormal_2009} and correspond to
 the intrinsic oscillation modes of generic fields in BH spacetime~\cite{coutant_dynamical_2016}.
Typically, localized modes have a finite lifetime due to radiative decay into waves propagating away from the horizon, for instance gravitational waves~\cite{chandrasekhar_quasi-normal_1975,berti_quasinormal_2009}.

While classical ring-down oscillations of a scalar field have been observed experimentallyfor surface waves in a rotating bathtub flow configuration~\cite{torres_quasinormal_2020} and a structured HE spectrum has been theoretically calculated in more complex flow geometries in conservative fluids~\cite{zapata_resonant_2011},
our work establishes that QNMs of quantum fields get naturally excited by the same quantum fluctuations that are responsible for the HE. 
Translated back to the astrophysical context, our results suggest a general and intrinsic fluctuations-driven mechanism leading to quantum excitation of BH spacetimes.

\begin{figure}[b!]
    \centering
    \includegraphics[width=\columnwidth]{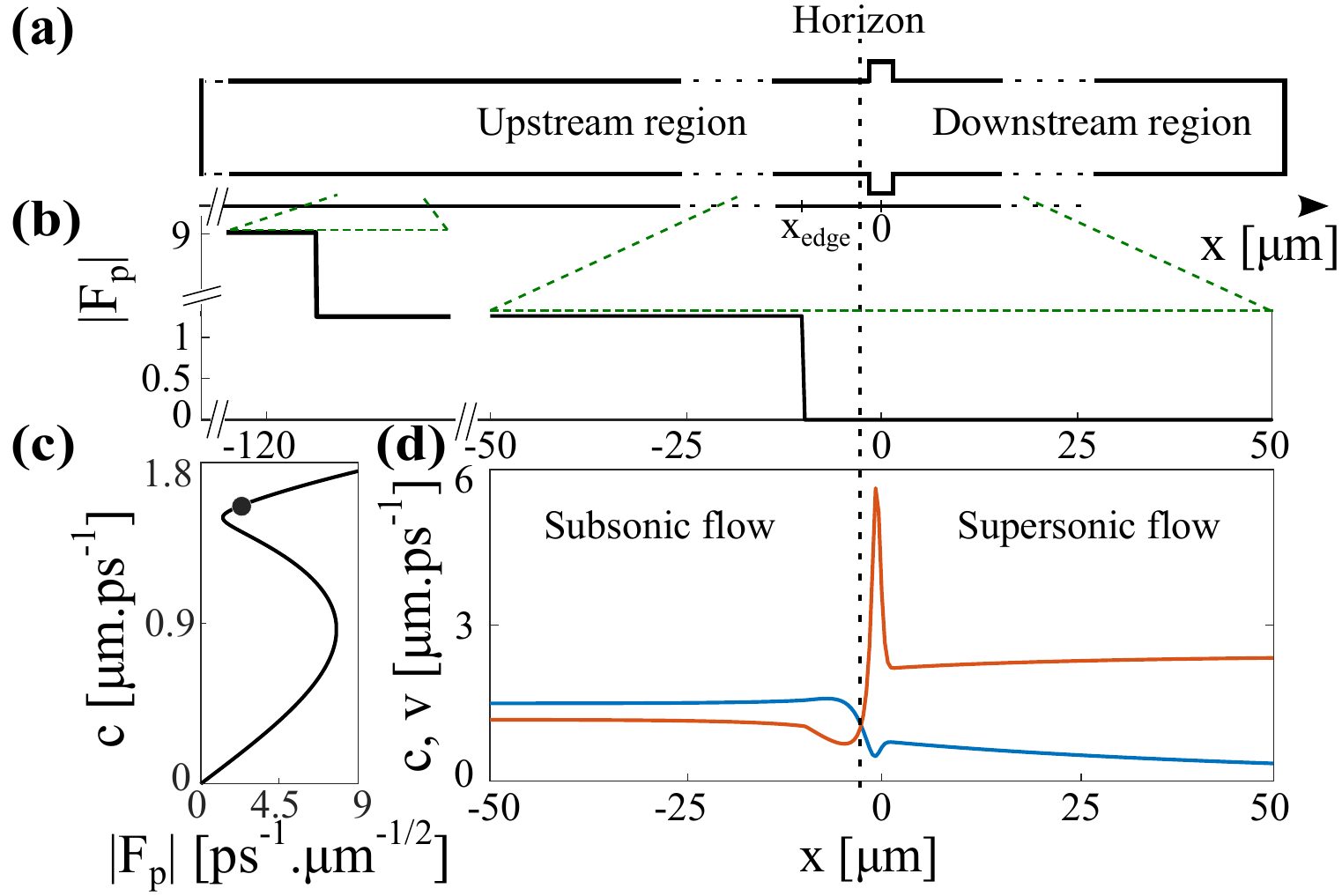}
        \caption{\textbf{Properties of the transsonic polariton fluid flow}.
    A step-like laser field pumps polaritons, creating a trans-sonic fluid flow across an attractive obstacle.
    \textbf{(a)} Sketch of the polariton wire with a defect at $x_d=0$.
    \textbf{(b)} Spatial profile of the pump near the defect. The pump intensity $|F_p|\, [\mathrm{ps}^{-1}.\mathrm{\mu m}^{-1/2}]$ drops abruptly to 0 at $x_{edge}=-\SI{10}{\micro\meter}$.
    \textbf{(c)} Optical bistability of the spatially homogeneous polariton fluid.
    \textbf{(d)} Spatial properties of the stationary fluid near the defect:
    orange, fluid velocity $v$; blue, speed of excitations $c$.}
    \label{fig:sys}
\end{figure}

\paragraph{Effective spacetime in a polariton fluid.--- }\label{sec:setup}
Our numerical study is based on the parameters of the system used in the experiment~\cite{Nguyen}: we consider a $\SI{800}{\micro\meter}$ long wire (a GaAs-based semiconductor microcavity embedding one InGaAs quantum well) within which the polariton dynamics is effectively one-dimensional.
There is an attractive defect formed by a $\SI{1}{\micro\meter}$ long broadening of the wire half-way along it at $x_d=0$, cf Fig.~\ref{fig:sys}~\textbf{(a)}.

The device is excited with a continuous wave laser (the pump) incident at a finite angle with respect to the normal to form a continuous flow of polaritons along the wire.
The pump frequency $\omega_p$ is tuned above the bare polariton resonance $\omega_0$ at the pump wavenumber $k_p$ to enforce a regime of optical bistability between the pump strength $|F_p|$ and the density of the fluid $n$~\cite{baas_bista_2004,carusotto_quantum_2013}, see Fig.~\ref{fig:sys}~\textbf{(d)} where we plot the speed of excitations $c=\sqrt{\hbar gn/m^*}$, as a proxy of the polariton density $n$ ($g$ is the interaction strength and $m^*$ is the polariton mass).
The pump spot is structured in a two-steps profile: on the first step, the fluid is set above the bistable regime     ($|F_p|$=\SI{9}{\per\pico\second\micro\meter\tothe{-1/2}}) on the second step the fluid density is supported near the turning point of the bistability loop ($|F_p|$=\SI{1.2}{\per\pico\second\micro\meter\tothe{-1/2}}, see black dot on Fig.\ref{fig:sys}~\textbf{(c)}) over about $\SI{100}{\micro\meter}$ up until the sharp pump edge $x_{edge}=\SI{-10}{\micro\meter}$ (cf Fig.\ref{fig:sys}~\textbf{(b)}).

In the region $x<0$, the phase $\theta$ of the fluid is locked to that of the pump, with a velocity $v = \frac{\hbar}{m^*}\partial_x \theta$ in the positive $x$ direction.
In the region $x>0$, polaritons propagate ballistically with an exponentially decaying density and a finite velocity~\cite{amelio_perspectives_2020}.
Fig.\ref{fig:sys}~\textbf{(d)} shows the fluid velocity $v$ in orange and the speed of excitations $c$ in blue.
The acoustic horizon separating the upstream sub-sonic ($v<c$) region from the downstream super-sonic ($v>c$) region is clearly visible at $x_H=\SI{-3}{\micro\meter}$.

\begin{figure}[ht!]
    \centering
    \includegraphics[width=.95\columnwidth]{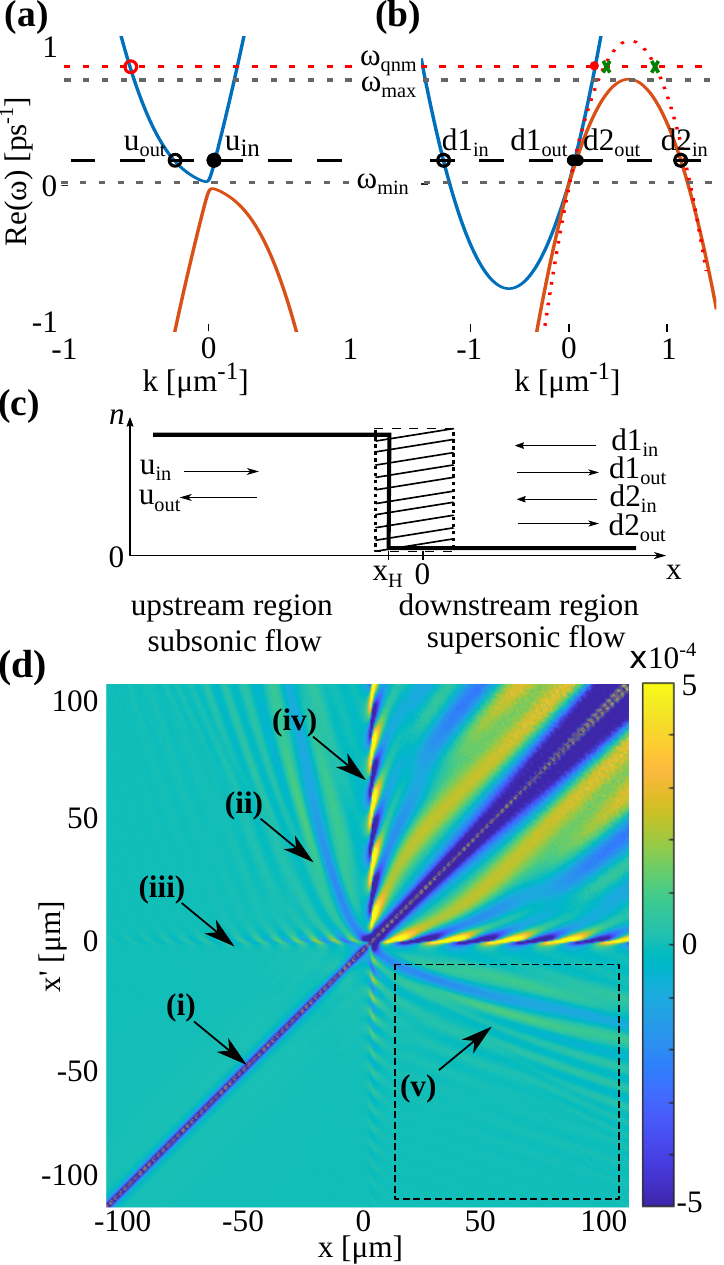}
    \caption{
    \textbf{Modes and correlated emission.}
    \textbf{(a)}-\textbf{(b)} Bogoliubov dispersion relation in the asymptotic regions as seen from the laboratory frame. \textbf{(a)} \textbf{Subsonic fluid flow}.
    \textbf{(b)} \textbf{Supersonic fluid flow}.
    Blue, positive-norm modes; orange, negative-norm modes; red-dashed, behavior of the negative-norm branch in the highly supersonic region.
    Circles (filled dots), asymptotic modes of negative (positive) group velocity.
    Gray dashed lines, $\omega_{min}$ and $\omega_{max}$; red dashed line, $\omega_{qnm}$; black dashed line a generic frequency giving to the mode structure of \textbf{(c)}.
    \textbf{(c)} \textbf{Modes of the system in the laboratory frame}. Schematic of the fluid density in the asymptotic regions (the dashed region has extra structure) and asymptotic mode structure on either side of the horizon.
    \textbf{(d)} \textbf{Normalized spatial correlations} $g^{(2)}(x,x')-1$. Traces: (i) antibunching; (ii) Hawking mustache; (iii) QNM and Hawking radiation; (iv) QNM and witness mode; (v) modulation of the Hawking mustache by QNMs.}
    \label{fig:corr}
\end{figure}

Excitations of a homogeneous fluid obey the $k$-dependent Bogoliubov dispersion relation $\omega=\pm\sqrt{\nicefrac{\hbar k^2}{2m^*}^2\left(\nicefrac{\hbar k^2}{2m^*}^2+2gn\right)}+vk-\nicefrac{i\gamma}{2}$, which depends on both $v$ and $n$ as well as on the particle loss rate $\gamma$.
It depends on the working point along the bistability loop and when driven, as in the upstream region, may be slightly gapped~\cite{carusotto_quantum_2013}.
Fig.~\ref{fig:corr} shows the dispersion in the asymptotic regions \textbf{(a)} up- and \textbf{(b)} downstream from the horizon.
The blue (orange) curve represents positive- (negative-)norm modes of the field, respectively~\cite{macher_black/white_2009}.
In the upstream region, the pump strength is tuned slightly above the turning point of the bistability loop to keep the system stable on the upper branch~\cite{jacquet_analogue_2022}: this introduces a small gap of size $\omega_{min}$ between the positive- and negative-norm modes in the upstream region.
As $v<c$, the Doppler shift is small and the positive- (negative-)norm modes remain at positive (negative) frequencies.
In the downstream region of ballistic flow, the dispersion recovers the gap-less Bogoliubov dispersion of conservative atomic condensates.
As $v>c$, the large Doppler effect pulls negative-norm modes to positive frequencies up to $\omega_{max}$.

As is typical in analogue systems based on quantum fluids, the description of collective modes in terms of an effective metric on a curved space-time is only valid for small $k$ modes~\cite{macher_black/white_2009,Recati_2009,corleyHawkingSpectrumHigh1996,jacquet_influence_2020}.
For larger wavenumbers, the superluminal form of the dispersion allows sound waves to propagate against the supersonic fluid flow in the downstream region.
The combination of the superluminal correction of the dispersion and the Doppler effect is, in fact, what enables the mixing of positive- and negative-norm modes at the horizon for frequencies within the interval $[\omega_{min},\omega_{max}]$.

\paragraph{Hawking effect.---}\label{sec:Hawking}This mode mixing at the horizon is the HE, i.e., the conversion of quantum fluctuations in the in-going $\{u_{in},d1_{in},d2_{in}\}$ modes propagating towards the horizon into pairs of real excitations in the out-going $\{u_{out},d1_{out},d2_{out}\}$ modes  (cf Fig.~\ref{fig:corr} \textbf{(c)})~\cite{Recati_2009,macher_black/white_2009}.
The correlated nature of the emission is visible in the intensity correlation diagram shown in Fig.~\ref{fig:corr} \textbf{(d)}, where the normalized correlation function $g^{(2)}(x,x')$ of density fluctuations (computed with the Truncated Wigner method~\cite{carusotto_parametric_threshold_2005} as a statistical average over $10^9$ Monte-Carlo realizations) is plotted as a function of the positions $x,x'$.
Here, besides the trivial antibunching (i) along the $x=x'$ diagonal stemming from repulsive interactions, the usual signature of HE is apparent as a $\SI{100}{\micro\meter}$-long oblique `mustache' of relative amplitude 
$\approx 10^{-3}$ (ii) due to correlations across the horizon between Hawking radiation ($u_{out}$) and the witness ($d1_{out}$) and partner ($d2_{out}$) modes in the downstream region.

\begin{figure}[t!]
    \centering
    \includegraphics[width=.8\columnwidth]{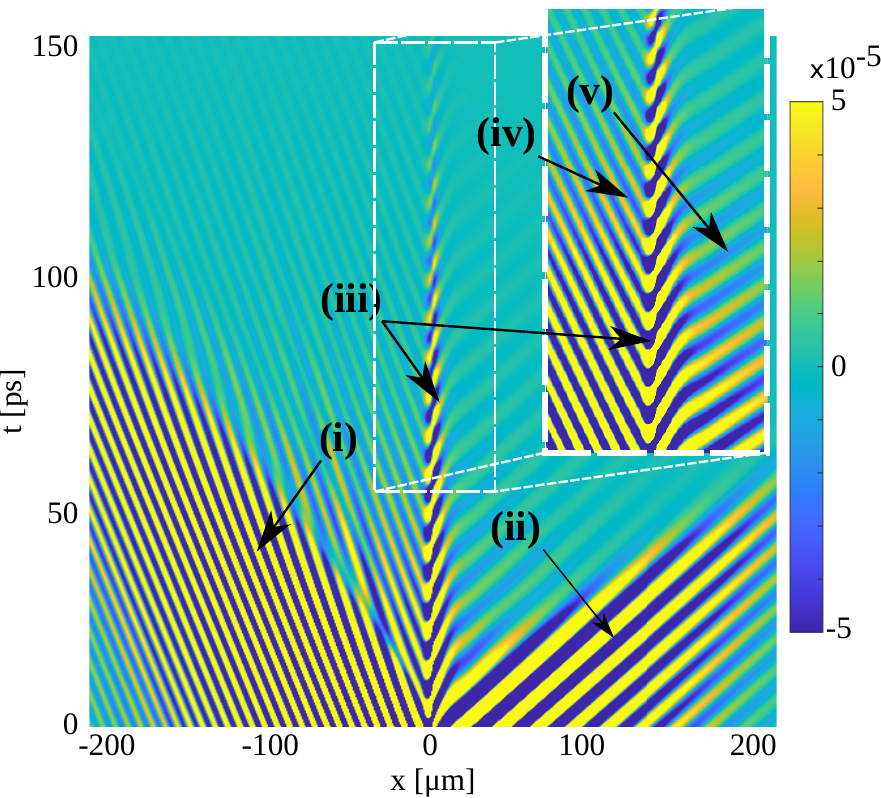}
    \caption{\textbf{Temporal evolution of the fluid density in response to an incident wavepacket}. $x_H=\SI{-3}{\micro\meter}$. Fluid density variation after excitation by a wavepacket ($\Delta k_{WP}= \SI{0.1}{\per\micro\meter}$, carrier frequency $\omega_{WP}=\omega_{qnm}$) hitting the horizon at $t=0$. Traces: (i) reflected wave, (ii) transmitted wave, (iii) state bound to the horizon, (iv) mode propagating upstream, (v) mode propagating downstream.}
    \label{fig:QNM}
\end{figure}

\paragraph{Localized correlation features.---}
In addition to these typical features of the HE, the correlation plots of Fig.~\ref{fig:corr} \textbf{(d)} reveal new structures: (iii) a series of oblique interference fringes developing along the $x=0,x'>0$ half 
line (and symmetrically along the $x'=0,x<0$ half line), and (iv) a similar, stronger series of oblique fringes localised along the $x=0,x'>0$ half line (and symmetrically along the $x'=0,x>0$ half line).
These traces indicate correlations between excitations in a mode spatially localized near the horizon and waves propagating in either the upstream (iv) or the downstream (iii) regions.
The spacing of the fringes yields the wavenumber of the propagating modes (see SM): $|k_u|=\SI{0.78\pm0.01}{\per\micro\meter}$ and $|k_d|=\SI{0.25\pm0.003}{\per\micro\meter}$ in the up- and downstream region, respectively.
Via the dispersion relation, $k_{u,d}$ correspond to the same frequency $\omega_{qnm}\sim \SI{0.75}{\per\pico\second}>\omega_{max}$ indicated by the red dashed line in Fig.~\ref{fig:corr}~\textbf{(a)-(b)}, and to modes $u_{out}$ and $d1_{out}$ that both have a positive norm.
This single frequency indicates that we observe a single mode of the acoustic field, composed of a localized component coupled to propagating waves on either side of the horizon.

\paragraph{Ringdown of the acoustic field.---}
The temporal dynamics of these coupled modes are not resolved by the equal-time correlation diagram Fig.~\ref{fig:corr}~\textbf{(d)} which displays the stationary properties of the emission.
To assess these dynamics, we simulate the temporal evolution of the acoustic field upon perturbation by a probe.
Fig.~\ref{fig:QNM} shows the temporal-evolution of the spatial density profile of the fluid in response to a generic wavepacket impinging on the horizon from the upstream $x<0$ region.
After the impinging wavepacket has been partially reflected (i) and transmitted (ii) by the horizon, a long-lived oscillation persists in the region near the horizon (iii): here, 
the acoustic field undergoes ring-down oscillations at the frequency $\omega_{qnm}$, with a lifetime $1/\gamma_{qnm}\approx\SI{14}{\pico\second}$ about the same as the bare polariton lifetime $1/\gamma$.
The coupling of the localized mode with the propagating (iv) $u_{out}$ and (v) $d1_{out}$ modes in the up- and downstream region, respectively, is also visible in Fig.~\ref{fig:QNM}.

This kind of ring-down is typical of the oscillation of quasi-normal modes (QNMs) of fields on curved spacetimes upon short perturbations~\cite{berti_quasinormal_2009}.
This oscillation occurs at the natural frequency of the underlying spacetime, independently of the details of the perturbation (as in BH merging for example~\cite{abbott_observation_2016}).
This is also the case here as the signatures of the ring-down are robust against variations in the properties of the impinging wavepacket,  e.g. its frequency $\omega_{WP}$ (there is a resonance when $\omega_{WP}=\omega_{qnm}$).
The real part of the frequency $\omega_{qnm}$ as well as the wavenumbers $k_{u,d}$ obtained here are the same as those extracted from features (iii) and (iv) in the correlation diagram of Fig.~\ref{fig:corr}~\textbf{(d)}, confirming that the same QNM is responsible for both the ring-down and these features.

\paragraph{Physical origin of the QNM.---}
As can be seen in Fig.~\ref{fig:sys}\textbf{(c)}, the fluid density displays a marked dip.
This dip is located in the inner region immediately after the horizon and is surrounded by two regions of different fluid density, forming an effective resonator for the Bogoliubov modes.
Because of  approximate conservation of the current, the density dip is accompanied by a narrow spike in the flow velocity $v$ leading to a narrow region of highly supersonic flow. This is responsible for a large Doppler shift of the Bogoliubov modes, which, in particular, pulls the negative-norm branch to highly positive frequencies above $\omega_{max}$ (red-dotted curve in Fig.~\ref{fig:corr}~\textbf{(b)}).
 Inside the resonator, negative-norm modes experience an effectively attractive potential~\cite{giacomelli_ergoregion_2020} that pins their wavenumber (green crosses in Fig.~\ref{fig:corr}~\textbf{(b)}), creating a standing-wave-shaped localized mode of frequency $\omega_{qnm}$.
This localized negative-norm mode tunnel-couples out into the propagating positive-norm modes $u_{out}$ and $d1_{out}$ in the asymptotic up- and downstream region, respectively (red circle and dot in Fig.~\ref{fig:corr}~\textbf{(b)}).
This is confirmed by a direct diagonalization of the inhomogeneous Bogoliubov problem, as discussed in the SM.

\paragraph{Quantum excitation of the QNM.---} This tunnel-coupling between modes of opposite norm-sign leads to the parametric amplification of vacuum fluctuations.
In our configuration, the interfaces of the narrow resonator inside the horizon have reflection coefficient effectively larger than unity for negative-norm waves: while this suggests that the field amplitude inside the resonator should grow exponentially, in our driven-dissipative fluid the instability is suppressed by the overall decay of the Bogoliubov modes due to the finite polariton lifetime~\cite{carusotto_quantum_2013}, making the system dynamically stable.
Similarly to parametric amplifiers driven below the instability threshold~\cite{tang_fundamentals_1995}, the parametric process is responsible for a sizable steady-state excitation of the QNM via the spontaneous parametric emission of paired excitations in the localized mode and in the propagating $u_{out}$ and $d1_{out}$ modes.

While the dynamical stability of the polariton fluid strongly depend on particle losses, the vacuum excitation of QNMs does not. In fact, the same phenomenon
can also be observed with a conservative fluid.
For example, in the SM we propose a toy-model configuration displaying an effective resonator near the horizon of a transsonic atomic BEC and show that such a system can be dynamically stable and display a QNM at $\omega_{qnm}<\omega_{max}$ that acquires a steady-state population with a clear signature in the correlation diagram of density fluctuations as well.
These results confirm that our predicted effect is generic to fluid-based analog models. While damping is the crucial ingredient for a steady-state quantum vacuum excitation, it may have different origins depending on the driven-dissipative (particle losses) or conservative nature of the fluid (radiative decay into propagative waves on either side of the horizon). 

\begin{figure}[ht!]
    \centering
    \includegraphics[width=.8\columnwidth]{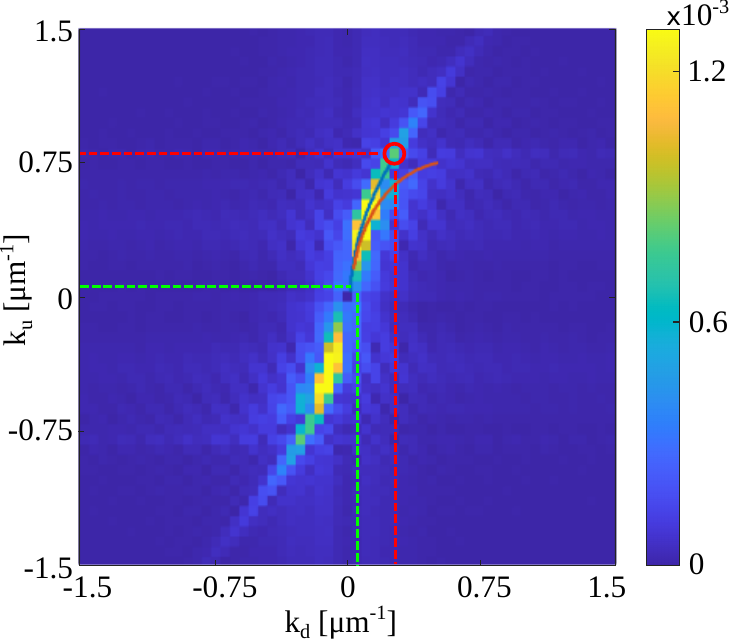}
    \caption{\textbf{Hawking emission spectrum}. Two-dimensional FT of $g^{(2)}(x,x')$ within the black rectangle in the up-downstream region in Fig.~\ref{fig:corr} \textbf{(d)}.
    Orange line, analytical locus of $u_{out}-d2_{out}$ correlations;
    Blue line, analytical locus of $u_{out}-d1_{out}$ correlations.
    Green dotted line, wavenumber cut-off $k_u=\SI{0.03\pm0.03}{\per\micro\meter}$ and $k_d=\SI{0.01\pm0.03}{\per\micro\meter}$ (due to the low-frequency gap in upstream dispersion, cf Fig.~\ref{fig:corr} \textbf{(a)}); Red dotted line, wavenumbers of the propagating component of the QNM $k_u=\SI{0.78\pm0.03}{\per\micro\meter}$ and $k_d=\SI{0.25\pm0.03}{\per\micro\meter}$.}
    \label{fig:FT}
\end{figure}

\paragraph{Spectral modulation of correlated emission.---}
We obtain the spectrum of correlated emission via a 2-dimensional Fourier Transform (FT) of the spatial correlations $g^{(2)}(x,x')$~\cite{steinhauer_measuring_2015,robertson_assessing_2017,isoard_departing_2020,isoard_bipartite_2021}.
The contribution of the horizon region is suppressed by restricting the FT to the sub-space delineated by the black dashed rectangle in the South-East quadrant in Fig.~\ref{fig:corr} \textbf{(d)}, where correlations between propagating modes in the up- and downstream regions (including the Hawking mustache) lay.
As shown in Fig.~\ref{fig:FT}, the FT of real data yields a pair of symmetric spectral lines centered in $k_{u,d}=0$.
These lines follow the locus of $k_{u,d}$ spanned by the dispersion of modes $u_{out}-d2_{out}$ (orange line) and $u_{out}-d1_{out}$ (blue line) as a function of $\omega$.
As anticipated in~\cite{boiron_quantum_2015,fabbri_momentum_2018}, this is the characteristic signature of the HE in $k$-space.
While they are well separated at large $k_{u,d}$, the tongues merge at low $k_{u,d}$ because of the limited size of the sampling box and the further spectral broadening due to the finite life-time of Bogoliubov excitations.
The spectral lines are cut off at small $k_{u,d}$ corresponding to the $\omega_{min}$ gap in the dispersion in the upstream region, and extend to large $k_{u,d}$ corresponding to $\omega_{max}$ in the downstream region.

Most importantly, beyond $\omega_{max}$, the emission spectrum shows an additional peak at $\SI{0.78\pm0.03}{\per\micro\meter}$, $\SI{0.25\pm0.03}{\per\micro\meter}$ in the $k_{u,d}$ plane.
Within the uncertainty due to the numerical grid size in Fig.~\ref{fig:FT}, these coordinates correspond to the wavenumbers extracted previously for the ringdown at frequency $\omega_{qnm}$.
This non-trivial structure of the spectrum is at the origin of the additional fringes next to the Hawking mustache indicated as feature (v) in Fig.~\ref{fig:corr}~\textbf{(d)}.
So this peak highlights a sizable additional contribution to the $u_{out}-d1_{out}$ correlations that are normally quite weak and indicates the spectral modulation of the HE by the quantum-fluctuation-driven excitation of the QNM.
A similar peak is visible in the HE spectrum in conservative fluids as well (see SM), although the peak there is located at a frequency $\omega_{qnm}<\omega_{max}$.
This indicates that the effective resonator acts as a frequency filter for the Hawking spectrum, i.e. a gray-body factor.

\paragraph{Conclusions.---}
In this Letter we have made use of an effectively curved spacetime realized in a polariton fluid to theoretically show that vacuum quantum fluctuations near a sonic horizon do not only yield the correlated emission of propagating waves by the Hawking effect but also cause a stationary quantum excitation of a quasi-normal mode of the acoustic field.
Specific features of this process are visible both in the correlation diagram of density fluctuations and in the spectrum of the Hawking emission.
Since our calculations are performed with experimental parameters and give a strong signal, the effects are amenable to experimental observation with state-of-the-art technology.
Although the microscopic origin of the decaying QNM mode may differ in driven-dissipative and conservative quantum fluids, similar signatures of the vacuum excitation of the QNM are visible in both cases. This suggests that our conclusions are valid for generic quantum fluids.

But the generality of the vacuum excitation of QNMs goes even beyond analog models.
Indeed, it originates in basic processes of quantum field theory in effectively curved spacetimes whose equivalence with Klein-Gordon fields on black-hole spacetimes is well established~\cite{Unruh,visser_acoustic_1998}.
For example, resonance peaks in the Hawking spectrum of astrophysical black holes appear via gray-body factors~\cite{page1976particle} and are associated with unstable orbits~\cite{goebel1972comments}.
In the quantum fluid configuration studied here, the resonator is located inside the horizon.
This is possible because of the specific superluminal dispersion of the fluid.
However this needs not be the case in general -- similar physics can be observed with resonators outside the horizon.
In astrophysical black holes, the localized component of the QNM is supported in the region between the light ring and the horizon and its amplitude naturally damps via radiative decay into propagating waves.
This suggests that our conclusions directly extend to the gravitational context, so that all quantum fields in curved spacetimes would exhibit the effects predicted in this work.
As such, the enhanced quantum fluctuations of the field in the localized component of QNMs raise questions on the role of their associated entropy~\cite{york_dynamical_1983,hod_bohrs_1998,maggiore_physical_2008} and on the intrinsic fluctuations of black-hole spacetimes.

\begin{acknowledgements}
We thank Th\'eo Torres, Riccardo Sturani, Luciano Vanzo, Stefano Vitale, and Mathieu Isoard for insightful discussions on QNMs and/or quantum fluids. We also thank Tangui Aladjidi for his careful reading of the manuscript. We acknowledge financial support from the H2020-FETFLAG-2018-2020 project ``PhoQuS'' (n.820392). IC and LG acknowledge financial support from the Provincia Autonoma di
Trento and from the Q@TN initiative. QG and AB are members of the Institut Universitaire de France.  
\end{acknowledgements}

%\bibliographystyle{apsrev4-2}
%\bibliography{bista-hawking_biblio.bib}

%apsrev4-2.bst 2019-01-14 (MD) hand-edited version of apsrev4-1.bst
%Control: key (0)
%Control: author (72) initials jnrlst
%Control: editor formatted (1) identically to author
%Control: production of article title (-1) disabled
%Control: page (0) single
%Control: year (1) truncated
%Control: production of eprint (0) enabled
%

\end{document}

\appendix
\newpage\onecolumngrid
\section{Supplemental material}
\subsection{Numerical methods}
Here we present the numerical method used to compute the quantum dynamics of the polariton fluid.
We refer the reader to ref~\cite{jacquet_analogue_2022}
for more details.
We use a quantum Monte-Carlo method called the truncated Wigner approximation (TWA): the equation of motion is truncated so as to map it to a stochastic partial differential equation for a classical field $\psi$:
\begin{equation}\label{eq:stoGPE}
    i \mathrm{d}\psi = \left[\omega_0 - \frac{\hbar}{2m^*}\frac{\mathrm{d}^22}{\mathrm{d}x^2} + V + g(|\psi|^2 - 1/\Delta x) - i\frac{\gamma}{2}\right]\psi \mathrm{d}t+ F_p \mathrm{d}t + \sqrt{\frac{\gamma}{4 \Delta x}} \mathrm{d}W, 
\end{equation}
where $\Delta x$ is the spatial grid and $\mathrm{d} W$ is complex white noise.
In numerical simulations, sampling of the realisations obtained with~\eqref{eq:stoGPE} starts when the steady state is reached (here we wait for 4ns).

Quantum observables are computed with statistical averaging over the realisations obtained with the TWA: the general rule for $N$ arbitrary observables is~\cite{carusotto_parametric_threshold_2005}
\begin{equation}\label{eq:twaobservables}
    \expval{O_1...O_N}_W 
    = \frac{1}{N!} \sum_{\tiny{\substack{\text{All} \\\text{$N$-permutations}}}} \expval{P(\hat O_1,...,\hat O_N)},
\end{equation}
where $\expval{}_W$ denotes the statistical averaging over the realisations.
\subsection{Parameters for the numerical calculation}
Parameters from~\cite{Nguyen}: $\hbar\gamma =\SI{0.047}{\milli\eV}$, $\hbar g =\SI{0.0003}{\milli\eV\micro\meter}$, $m^* = 3 \cdot 10^{-5} m_e$. We use a detuning between the pump frequency and polariton frequency (at $k=0$): $\omega_p-\omega_0 = \SI{0.49}{meV}$. Cavity length $\SI{800}{\micro\meter}$. Grid size 2048 points.
\subsection{Wavenumber of the propagating components of the QNM}

\begin{figure}[h!]
    \centering
    \includegraphics[width=.4\columnwidth]{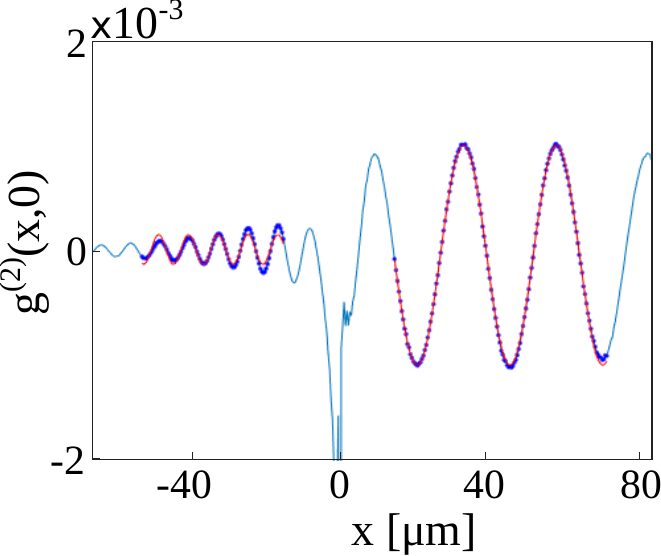}
    \caption{\textbf{Density correlations along the $x'=0$ line in Fig.~2}.
    Fit (purple) of the correlations (blue).}
    \label{fig:kQNM}
\end{figure}

The wavenumber of the propagating components of the QNM in the upstream and downstream region is extracted by fitting the interference fringes along the $x'=0$ line in Fig.~2~\textbf{(d)}.
From the fit in Fig.~\ref{fig:kQNM}, at $x<0$ we obtain $|k_u|=\SI{0.78\pm0.01}{\per\micro\meter}$, while at $x>0$ we obtain $|k_d|=\SI{0.25\pm0.003}{\per\micro\meter}$.

\subsection{Fit of the ringdown of the density in the horizon region}
\begin{figure}[h!]
    \centering
    \includegraphics[width=.4\columnwidth]{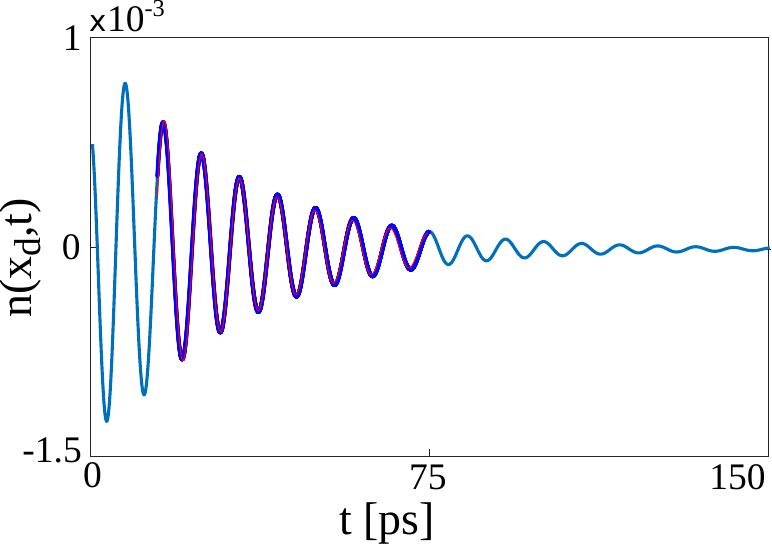}
    \caption{\textbf{Time evolution of the density $n(x,t)$ at $x_d=\SI{0}{\micro\meter}$ in Fig.~3}.
    Fit (purple) of the decay of the density oscillations (blue). The mean density of the fluid has been substracted.}
    \label{fig:RD}
\end{figure}
The complex frequency of the QNM is obtained by fitting the decay of the oscillations at $x_d=\SI{0}{\micro\meter}$ in Fig.~3 after a generic wavepacket has impinged upon the horizon from the upstream region.

In Fig.~3 the wavepacket has a width $\Delta k=\SI{0.1}{\per\micro\meter}$ and a duration of $\SI{58}{\pico\second}$.
Its front impinges on the horizon at $t=\SI{-58}{\pico\second}$ and its back finishes interacting with the horizon until $t\approx0$, after which the density at $x_d$ undergoes ringdown.
In Fig.~\ref{fig:RD} the fluid density is sampled from  $t=0$.
The ringdown oscillations are fitted with an exponential giving the complex frequency of the QNM: $\mathrm{Re}(\omega_{qnm})=\SI{0.75}{\per\pico\second}$ and $\mathrm{Im}(\omega_{qnm})=\SI{0.036}{\per\pico\second}$.

\subsection{Diagonalization of the inhomogeneous Bogoliubov problem}

Further confirmation and characterization of the physics of the QNM we identified can be obtained by numerically diagonalizing the Bogoliubov problem describing linear perturbations around the stationary state. In particular, once the steady state $\Psi_{ss}$ of the polariton fluid in the cavity has been obtained by evolving for a long enough time the generalized Gross--Pitaevskii equation (GPE)
\begin{equation}
    i\partial_t\Psi_{LP}=\left[\omega_{LP}^0-\frac{\hbar\nabla^2}{2m^*}+V+g|\Psi_{LP}|^2-i\frac{\gamma}{2}\right]\Psi_{LP}+iF_p=:\left[H_{GP}-i\frac{\gamma}{2}\right]\Psi_{LP}+iF_p
    \label{eq:ddgpe}
\end{equation}
one can consider small fluctuations $\Psi_{ss}+\delta \Psi$.

The equations governing the fluctuations can be obtained by linearizing the GPE and can conveniently be written in terms of the spinor $(\delta \Psi, \delta \Psi^*)^T$
\begin{equation}\label{eq:bogo-problem}
	i\partial_t
	\begin{pmatrix}
		\delta\Psi\\ \delta\Psi^*
	\end{pmatrix}
	=
	\begin{bmatrix}
		H_{GP}+g|\Psi_{ss}|^2-i\gamma/2 & g\Psi_{ss}^2 \\
		-g\left(\Psi_{ss}^*\right)^2 & -H_{GP}-g|\Psi_{ss}|^2-i\gamma/2
	\end{bmatrix}
	\begin{pmatrix}
		\delta\Psi\\ \delta\Psi^*
	\end{pmatrix}.
\end{equation}

This Bogoliubov matrix can be represented numerically on a spatial grid of $N$ points and by using discretized expressions for the derivatives. The resulting $2N\times 2N$ matrix can then be diagonalized to obtain the eigenvalues and the corresponding eigenmodes.

The eigenmodes $\ket{\phi_i}=(u_i,v_i)^T$ with frequencies $\omega_i$ of the Bogoliubov problem can be distinguished by the sign of their \textit{norm} $\Vert\phi_i\Vert=\int dx (|u_i|^2-|v_i|^2)$. An eigenmode $\ket{\phi_i}$ corresponds to a fluctuation on the fluid wavefunction
\begin{equation}
    \delta\Psi_i(x,t)=u_i(x)e^{-i\omega_i t}+v_i^*(x)e^{i\omega_i t},
\end{equation}
so that the $v$ component is the negative frequency component of the excitation mode. Given that the energy of an eigenmode is $E_i=\Vert\phi_i\Vert\hbar\omega_i$, at positive frequencies negative-norm modes have a negative energy, and, more in general, the spatial regions where $|v_i(x)|>|u_i(x)|$ contribute with a negative energy to the mode.

Because of the dissipative terms appearing on the diagonal of~\eqref{eq:bogo-problem}, all eigenmodes will decay with the polariton lifetime. An exception are zero-norm modes, that carry zero energy and whose imaginary part deviates from $-\gamma/2$. These can emerge from the resonance of modes with opposite norms.
In conservative fluids, this leads to dynamical instabilities, i.e. to the exponential growth of fluctuations.
See Ref.~\cite{castin2001bose} for a complete discussion of the properties of the Bogoliubov problem.

\begin{figure}[t]
    \centering
    \includegraphics[width=\columnwidth]{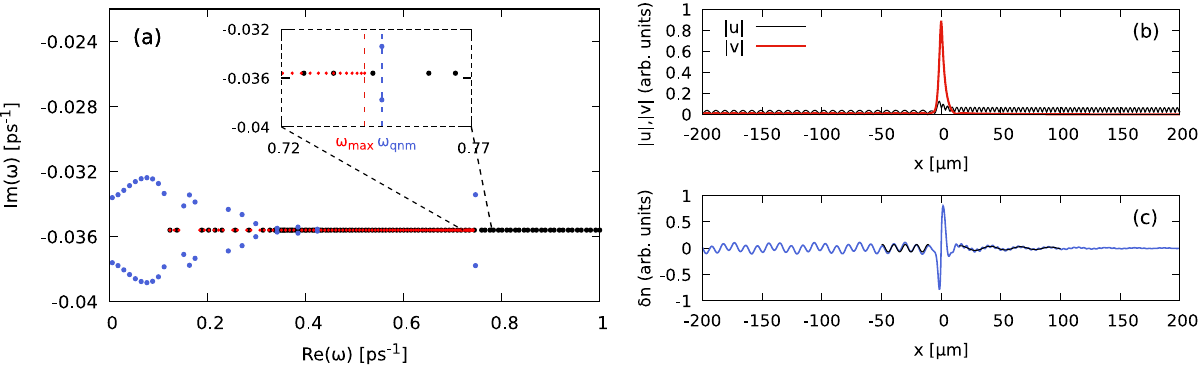}
    \caption{\textbf{Result of the diagonalization of the inhomogeneous Bogoliubov problem.} \textbf{(a)} Spectrum of linear excitations, with black (red) dots indicating positive-(negative-)norm modes and blue ones indicating zero-norm modes. The zero-norm mode just above $\omega_{max}$ is the one responsible for the ringdown of the acoustic field (magnified spectrum in the inset). The other zero-norm modes are spurious effects due to the boundary conditions. \textbf{(b)} Modulus of the components of the Bogoliubov spinor of the QNM; black line, $u$ component; red line, $v$ component. \textbf{(c)} Corresponding density fluctuations. Besides high frequency interference fringes, oscillations at the same wavelength as those observed in Fig.~\ref{fig:kQNM} (black fit) are visible. %As a comparison, sines with these expected wavelengths are plotted as black lines on top of the numerical data.
    }
    \label{fig:diagonalization}
\end{figure}

In the numerical calculations reported in the main text, absorbing potentials at the edges of the numerical box are used to mimic the experimentally relevant situation of an open system at the extremities of the wire. However these non-conservative boundary conditions cannot be included in the diagonalization so we consider periodic boundary conditions with a large enough numerical range to fit the decay of the fluid in the downstream region, see Fig.~\ref{fig:diagonalization}.

Fig.~\ref{fig:diagonalization} \textbf{(a)} shows the positive frequency spectrum of positive-norm modes (black dots), negative-norm modes (red dotes) and zero-norm modes (blue dots).
As usual in Bogoliubov theories, the spectrum at negative real frequencies is specular, with positive- and negative-norm modes interchanged.
The imaginary part of the zero-norm modes spectrum departs from the $-\gamma/2$ common to the other modes. 
In agreement with the dispersion plotted in 
Fig.~2~\textbf{(a)} of the main text, negative-energy modes are present up to the frequency $\omega_{max}$.
All zero-norm modes that are visible below this frequency are a spurious consequence of the periodic boundary conditions, which induce black-hole lasing effect around the integration box.
These zero-norm modes do not exist in the open-boundary system considered in the main text as they emerge when negative-norm states are fed back to the horizon on the other side after traveling around the integration box.
This interpretation is confirmed by the fact that the real part of the frequencies of all these zero-norm modes changes with increasing size of the numerical box, while their imaginary part approaches the value $-\gamma/2$ of positive/negative norm modes.
On the other hand, the real part of the zero-norm mode just above $\omega_{max}$ is $\omega_{qnm}$ (as measured from the wavelength of the correlation features in Fig.~\ref{fig:kQNM} and from the ringdown numerical experiment in Fig.~\ref{fig:RD}) for any size of the numerical box, which points at its physical nature as localized quasi-normal mode.

The QNM frequency $\omega_{qnm}$ does not depend on the system size because it is determined by the part of the mode localized near the defect.
Localization can be seen in Fig.~\ref{fig:diagonalization}~\textbf{(b)} and \textbf{(c)}, which displays the modulus of the components of this Bogoliubov spinor and the corresponding density fluctuations.
The mode has a sharply peaked negative-energy part around $x=0$, while it is composed of positive-energy waves on either side of the horizon.
Again, interference fringes there are a spurious effect due to the periodic boundary conditions.

These results evidence the existence of a mode with the features deduced from the correlation diagram Fig.~3.
The QNM thus originates from trapped negative-energy standing waves in the horizon region.
In the present case, only one QNM exists, but several peaks at distinct frequencies exist for larger defect widths, giving a discrete QNM spectrum of the acoustic field on the effective black hole spacetime.

\subsection{Excitation of a QNM in a conservative fluid}

\begin{figure}[ht]
    \centering
    \includegraphics[width=.5\columnwidth]{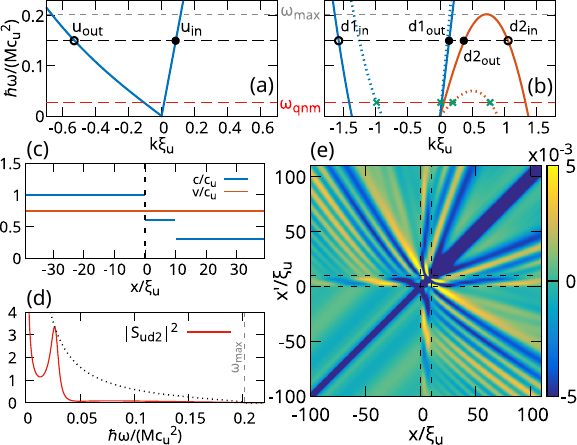}
    \caption{\textbf{QNM features of the Hawking effect in an atomic-fluid-based analog model.} \textbf{(a)-(b)} Dispersion relation in (a) the subsonic region and (b) the resonator (dashed line) and asymptotic supersonic (solid lines) regions. Blue, positive-norm modes; orange, negative-norm modes. \textbf{(c)} Spatial dependence of the fluid flow velocity (orange) and of the speed of sound (blue). \textbf{(d)} Hawking spectrum in the configuration of panel (c). \textbf{(e)} Corresponding $g^{(2)}(x,x')-1$. The units in all panels are given in terms of the atomic mass $M$, the speed of sound in the first region $c_u$ and the healing length $\xi_u=\hbar/Mc_u$.}
    \label{fig:BEC}
\end{figure}

While the specific profile of the polariton fluid considered in the main text crucially depends on the presence of driving and particle losses, the phenomenon of the vacuum excitation of QNMs does not.

In the polariton configuration, the presence of the QNM originates from a resonator just inside the horizon, within which the density dips and the flow velocity is supersonic.
The resonator appears because of the impedance mismatch between this narrow region and the asymptotic regions outside and inside the horizon: outside the horizon the density is high and the flow velocity is low, while inside the horizon the density is low and the flow velocity is high.

The fact that the flow in the resonator region is more supersonic than in the inside and outside region implies that there is a range of frequencies above $\omega_{max}$ in which negative-energy modes are available only inside the resonator (see Fig.1\textbf{(a)}-\textbf{(b)} of main text). As a result, because of energy conservation, the reflectivity of the two edges of the effective resonator for these modes are greater than unity.
A reflectivity larger than unity would normally lead to an exponential growth of the field amplitude inside the resonator, i.e. to a dynamical instability of the fluid.
But in our case this growth is suppressed by the limited lifetime of polaritons.

In contrast, a conservative fluid (e.g. a superfluid of ultracold atoms) featuring such a high-Q resonator would not be dynamically stable.
In some configurations, this would lead to the so-called black-hole lasing effect~\cite{leonhardt_theory_2003,barcelo_stability_2006,finazzi_black_2010,zapata_resonant_2011,giacomelli_ergoregion_2020}.
Beyond this, in what follows, we will show that it is possible to engineer configurations in which a conservative fluid does not become dynamically unstable even in the absence of particle losses.
To this end, we consider a toy-model configuration for a conservative atomic condensate fluid, that is described by the conservative version of the Gross--Pitaevskii equation~\eqref{eq:ddgpe} (i.e. without the loss and driving terms).

We take inspiration from refs.~\cite{zapata_resonant_2011,Giacomelli_superradianceBEC_2021} and extend the model of~\cite{carusotto_numerical_2008} to the three-region configuration sketched in Fig.~\ref{fig:BEC}(c): we consider a one-dimensional, spatially infinite condensate with constant flow velocity $v$ and an inhomogeneous speed of sound.
The fluid has constant density so we control the speed of sound via the interaction constant $g(x)$, which we tune by means of an external potential such that $V_\mathrm{ext}(x)+g(x)n=\mathrm{const}$~\cite{carusotto_numerical_2008}.
Finally, the transsonic fluid is built by assembling regions via step-like transitions: as a concrete example, we consider the configuration Fig.~\ref{fig:BEC}(c) that consists of a subsonic region (outside the horizon) and two supersonic regions (inside the horizon).
The supersonic region immediately inside the horizon has a larger speed of sound than the asymptotic region.
With this configuration, we create an effective resonator in the intermediate region of supersonic flow. Although this configuration was adopted in analogy with the polariton flow considered in the main text, effective resonators located at different spatial positions are obtained with different choices for the spatial profile of the speed of sound, yet with no qualitative change in the basic features of the Hawking emission spectrum and of the QNM fluctuations.

Fig.~\ref{fig:BEC} shows the dispersion relation in each of the three regions: the dispersion in subsonic region (a) is identical to the polaritonic configuration up to the gap near $k=0$).
On the other hand, the dispersion inside the horizon (b) differs from the polaritonic configuration.
The solid line shows the dispersion in the asymptotic region, that is qualitatively identical to the polaritonic case, while the dashed line shows the dispersion inside the resonator.
Also in this region the fluid flow is supersonic, but less strongly so than in the asymptotic region because of the higher speed of sound.
Therefore, the negative-norm branch is also pulled up to positive frequencies, but not as much as in the asymptotic region.
Thus, while the impedance mismatch between the subsonic region and the resonator is qualitatively equivalent to the one in the polaritonic configuration, there is no frequency interval in which negative-energy modes are available only in the resonator region.

We will now calculate the emission spectrum and the spatial correlations and show that this flow configuration is dynamically stable, implying that the reflection coefficient at the resonator/asymptotic interface is lower than unity.
In brief, it is this interface that provides the dissipation that suppresses the dynamical instability that would otherwise build up inside the resonator, and thus makes the conservative fluid stable.

The step-like change in fluid properties at each interface allow for the scattering problem to be solved semi-analytically by imposing continuity conditions at the interfaces~\cite{Recati_2009}.
This yields the total scattering matrix $S$ that encodes all statistics of the Hawking effect~\cite{jacquet_analytical_2020}.
The Hawking spectrum is directly given by the scattering coefficient $|S_{ud2}|^2$ for an incoming negative-norm mode $d2_{in}$ the outgoing positive-norm mode $u_{out}$ (Hawking radiation).
Fig.\ref{fig:BEC}(d) shows the Hawking spectrum (solid red line), that monotonously drops from $\omega=0$ until it cuts at $\omega_{max}\simeq0.20Mc_u^2$, but for a peak at $\hbar\omega_{qnm}\simeq0.03Mc_u^2$.
This spectrum largely departs from the monotonous $\approx1/\omega$ decay (gray dashed line) that is obtained for a configuration with the same inside/outside regions but without the resonator~\cite{Recati_2009,isoard_departing_2020}: while the two spectra span the same frequency range up to $\omega_{max}$, the spectrum with a resonator has lower amplitude except at $\omega_{qnm}$.
The resonator thus acts exactly like a Fabry-Perot resonator, i.e. a frequency filter that modifies the Hawking spectrum with a resonance peak.
Just like a gray-body factor.

Note that this is different from the polaritonic configuration in which the spectrum peaks at $\omega>\omega_{max}$.
In fact in that case we saw that the QNM is a zero-norm mode that emerges from the reflectivity larger than one at both edges of the resonator and whose decay is given by the finite lifetime of excitations. Also in the present case at the QNM frequency the reflectivity of the outer interface of the resonator is larger than one for negative energy waves. Instead, the inner interface has a reflectivity smaller than one because negative-energy modes are available in the inner region. Hence, the damping mechanism for the localized mode is in this case provided by open boundary conditions \cite{coutant_dynamical_2016}.

Finally, from the scattering matrix we compute (full details of the calculation technique to appear) the spatial correlations $g^{(2)}(x,x')$ shown in Fig.\ref{fig:BEC}(e).
The map is qualitatively similar to Fig. 2 (d) of the main text~\footnote{The difference in relative intensity of the features with respect to Fig. 2 of the main text is due to the non-constant density in that case, as we consider a normalized correlation function.}: the diagonal shows the fluid's autocorrelation, the fringes located along the horizontal and vertical axis signal the correlation of the resonator with the entire fluid up- and down-stream, and the Hawking moustache features a low-frequency transverse modulation corresponding to the emission peak at $\omega_{qnm}$.
Besides these new features, we also remark a high-frequency modulation that is due to dispersion in the system~\cite{isoard_departing_2020}.

To conclude, the microscopic origin of the QNM is the only notable difference between the conservative and polaritonic configurations.
In all cases, the existence of a decaying mode of the acoustic field implies that the field acquires a steady-state population inside the resonator that is correlated to propagating modes on both sides of the horizon.
This population is driven by the Hawking effect, whose spectrum is in turn modified by the QNM.
The microscopic origin of the decaying mode has no influence on the generality of the physics at play, be it in an experimentally relevant configuration like the polaritonic one or a toy-model like the one discussed in this SM.

This physics only relies on damping for a steady-state quantum vacuum excitation.
Thus it should be general to all analog models as well as to quantum fields on astrophysical black hole spacetimes.